\documentclass[twocolumn,showpacs]{revtex4}

\usepackage{graphicx}%
\usepackage{dcolumn}
\usepackage{amsmath}

\makeatletter
\def\btt#1{\texttt{\@backslashchar#1}}%
\DeclareRobustCommand\bblash{\btt{\@backslashchar}}%
\makeatother

\begin{document}


\title[Short Title]{Numerical Study of the Spin-Flop Transition in 
Anisotropic Spin-1/2 Antiferromagnets}

\author{Seiji Yunoki}
\affiliation{
Solid State Physics Laboratory, Materials Science Center, \\ 
University of Groningen, Nijenborgh 4, 9747 AG Groningen, The Netherlands
}%

\date{\today}

\begin{abstract}
Magnetization processes of the spin-1/2 antiferromagnetic $XXZ$ model 
in two and 
three spatial dimensions are studied using quantum Monte Carlo method based 
on stochastic series expansions. Recently developed operator-loop algorithm 
enables us to show a clear evidence of the first-order phase transition 
in the presence of an external magnetic field.
Phase diagrams of closely related systems, hard core 
bosons with nearest-neighbor repulsions, are also discussed focusing 
on possibilities of phase-separated and supersolid phases.  
\end{abstract}

\pacs{75.10.-b, 75.10.Jm, 74.20.Mn}

\maketitle


There have been a lot of interests on magnetic properties of anisotropic 
quantum antiferromagnets since {N\'eel} first predicted the first-order phase 
transition in the presence of an external magnetic field \cite{neel}. 
One of the simplest models for anisotropic antiferromagnets in an external 
magnetic field is described by the following spin-1/2 $XXZ$ model Hamiltonian: 
\begin{equation}
H = J \sum_{\langle{\bf i},{\bf j}\rangle}\left(S_{\bf i}^xS_{\bf j}^x 
+ S_{\bf i}^yS_{\bf j}^y + \Delta S_{\bf i}^zS_{\bf j}^z \right) 
- h\sum_{\bf i} S_{\bf i}^z, 
\label{xxz}
\end{equation}
where $S_{\bf i}^\alpha$ is the $\alpha$$(=x,y,z)$ component of 
the spin-1/2 operator at site {\bf i}, $h$ is an external magnetic field 
applied in the z-direction, and $\langle{\bf i},{\bf j}\rangle$ runs over 
all the nearest-neighbor pairs of spins at sites {\bf i} and {\bf j}.
$J(>0)$ is antiferromagnetic coupling and $\Delta(\ge0)$ is anisotropic 
constant. 
Mean-field calculations of the spin-1/2 $XXZ$ model~\cite{mf2}, 
supporting N\'eel's prediction, found the first-order phase transition 
from the N\'eel ordered state to the spin-flopping state with increasing 
the magnetic field $h$.  
On the contrary to these studies, it is known from Bethe ansatz solution 
that the one-dimensional (1D) 
spin-1/2 $XXZ$ model shows a second-order transition in the presence of 
the external magnetic field~\cite{yang}. The discrepancy is assigned to the 
inadequacy of treating quantum fluctuations by the mean-field theories. 
It is thus important to use an unbiased numerical method for understanding 
the correct nature of the magnetization process even for the simplest 
systems such as one given by Eq.~(\ref{xxz}) in higher spatial dimensions 
since there exist no analitically exact solutions. 
This is preciely one of our purposes for this paper~\cite{takahashi}. 

Another importance of studing the spin-1/2 $XXZ$ model defined by 
Eq.~(\ref{xxz}) comes from the fact that the model is 
mapped onto a system of hard core bosons with nearest-neighbor repulsions 
described by the Hamiltonian
\begin{equation}
H_{\rm B} = - t \sum_{\langle{\bf i},{\bf j}\rangle}
\left(c_{\bf i}^\dagger c_{\bf j} + c_{\bf j}^\dagger c_{\bf i}\right) 
+V\sum_{\langle{\bf i},{\bf j}\rangle} n_{\bf i} n_{\bf j} 
- \mu\sum_{\bf i} n_{\bf i}, 
\label{boson}
\end{equation}
with $t=J/2$, $V=J\Delta$, and $\mu=h+zJ\Delta/2$ 
($z$: coordination number)\cite{matsubara}. 
Here $c_{\bf i}^\dagger$ is a creation operator 
of hard core boson at site ${\bf i}$ and 
$n_{\bf i}=c_{\bf i}^\dagger c_{\bf i}$. The total magnetization 
$M_z = \sum_{\bf i} S_{\bf i}^z$ 
thus relates to $\sum_{\bf i}\left(n_{\bf i}-1/2\right)$ in the boson 
model. The boson Hamiltonian $H_{\rm B}$ is proposed as a model 
Hamiltonian to study properties of liquid $^4{\rm He}$~\cite{fisher}, 
granular superconducting arrays~\cite{cha}, and flux lines in 
superconductors~\cite{nelson}. The ground state phase diagram has been 
studied using mean-field theories~\cite{mf2,mf} and numerical 
methods~\cite{bat2}, 
and shows a Mott insulating phase, a superfluid phase, 
and a phase with having both the orders simultaneously 
($supersolid$ phase). The correspondence of these states to those in the 
spin model is as follows: Mott insulating and N\'eel states, 
superfluid and spin-flopping states, and supersolid and 
``intermediate'' spin states~\cite{mf2}, respectively.

The main purpose of this paper is to show a clear evidence for the 
first-order phase transition of the 2D and 3D spin-1/2 
$XXZ$ models in the presence of a magnetic field by using recently 
developed numerical method and present the ground state phase diagrams. 
The presence of a phase-separated phase and the absence of the 
supersolid phase in the closely related system of hard core bosons 
with nearest-neighbor repulsions are also discussed.

The magnetization process of the 2D and 3D spin-1/2 $XXZ$ models defined by 
Eq.~(\ref{xxz}) is studied 
numerically on square (number of spins $N_s=L\times L$) and cubic 
($N_s=L\times L\times L$) lattices using quantum Monte Carlo (QMC) 
technique based on stochastic series expansions (SSE)~\cite{sandvik}. 
Very recently an important technical improvement have been achieved 
by Sandvik~\cite{loop}. He found a new algorithm of cluster-type updates 
(operator-loop updates) within the SSE QMC scheme which reduces 
autocorrelation time drastically compared to simulations using only local 
updates. While this method is very similar to the loop algorithm in the 
world-line QMC method proposed by Evertz {\it et al.}~\cite{evertz},
one major advantage of the SSE method with operator-loop updates 
is that there is no difficulty in simulating systems with anisotropic 
couplings in external magnetic fields owing to not having to have 
``freezing'' configurations and ``global'' weights which make the loop 
algorithm in the world-line QMC method highly inefficient~\cite{loop}. 
This reduced autocorrelation time enables us to go down to very 
low temperatures in very high magnetic fields and therefore the method is 
suitable for our purpose. 
Another advantage of this study using the SSE scheme over other earlier 
numerical 
studies~\cite{takahashi,bat2,bat} is that the simulations are 
performed directly in the ground canonical ensemble, {\it i.e.}, 
magnetization per site $m_z=M_z/N_s$ is calculated for a given magnetic 
field $h$. 
In this paper temperatures $T$ are set to be $J/k_{\rm B}T=2L$ which 
is low enough to study the ground state properties on finite 
lattices~\cite{temp}, and periodic boundary condition is used. 
The exchange coupling $J$ will be taken as the energy unit.
Since there exists long-range {N\'eel}-ordered 
state only for $\Delta\ge1.0$, from which the first-order spin-flop 
transition can take place by applying a finite external field, 
our main focus in this paper will be put on this anisotropic 
regime.

\begin{figure}[thbp]
\includegraphics[clip=true,width=0.34\textwidth,angle=-90]{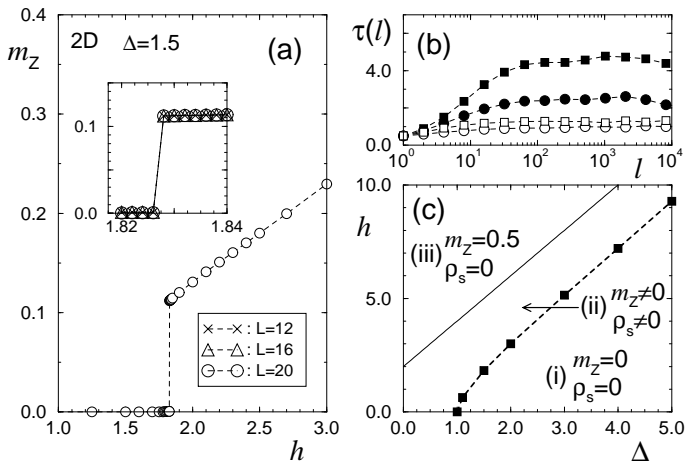}
\caption{(a) Magnetization curve of the 2D spin-1/2 $XXZ$ model with 
$\Delta=1.5$. Inset: enlarged scale is used. 
Error bars are smaller than symbols. 
(b) Correlation time $\tau(l)$ (for definition see in the text) 
of $m_z$ (circles) and $S(\pi,\pi)$ (squares) as a function 
of bin length $l$. The parameters are $N_s=12^2$, $T=1/24$, $\Delta=1.5$, 
and $h=2.0$. Solid (open) marks are data for operator-loop updates 
with closed loops constructed as many as 25 (100) times per MC step. 
(c) The ground state phase diagram of the 2D spin-1/2 $XXZ$ model with 
anisotropic constant $\Delta$ in the presence of the magnetic field $h$. 
There exist three phases: (i) N\'eel ordered phase, (ii) spin-flopping 
phase, and (iii) fully saturated ferromagnetic phase. 
Solid line is $2(1+\Delta)$ (see in the text). Dashed line is a guide to 
the eye. 
}
\label{2d}
\end{figure}

Let us first study the magnetization process in 2D. A typical example of 
the magnetization curves is shown in Fig.~\ref{2d} (a). One can see 
that at certain critical magnetic field $h_c$ the magnetization $m_z$ changes 
discontinuously from 0 to a finite value $m_z^c$. For this example in the 
figure with $\Delta=1.5$ the magnetization jumps at 
$h_c\sim1.83$ to $m_z^c\sim0.11$~\cite{boson}. Working on various system 
sizes (see in the inset of Fig.~\ref{2d}(a)), we confirm that finite size 
effects are small and conclude that the jump in magnetization is not due 
to extrinsic factors of working on finite-size lattices. 
The result already gives a clear evidence for the first-order phase 
transition.

In order to illustrate that autocorrelation times of our QMC measurements 
are short enough, the integrated autocorrelation time $\tau_{\rm int}$ 
is estimated as follows: we first divide a 
sequence of Monte Carlo data points into bins of length $l$, and 
for each bin length $l$, the average of the data in the $b$th bin and 
the variance $\sigma^2(l)$ of the bin averages are calculated. 
The integrated autocorrelation time $\tau_{\rm int}$ is then estimated by 
the asymptotic value of $\tau(l) = l\sigma^2(l)/2\sigma^2(l=1)$ at 
large $l$ above which $\tau(l)$ does not depend on 
$l$~\cite{kawa}. The results of $\tau_{\rm int}$ for $m_z$ and spin 
structure factor $S({\bf q})$ at ${\bf q}=(\pi,\pi)$ are shown in 
Fig.~\ref{2d} (b) for $\Delta=1.5$ and $h=2.0$. From this figure 
$\tau_{\rm int}$'s are estimated to be no longer than 4--5 MC steps. 
Since we use bin length $l=1000$ or more to estimate the statistical errors, 
we can be sure that our error bar is accurate. 
It should be noted here that $\tau_{\rm int}$ can be controlled by changing 
how many times closed loops are constructed in the operator-loop updates. 
In this paper we have to construct closed loops as many as about 1000 
times per MC step for larger values of $\Delta$.

Repeating the same procedure for different values of $\Delta$, the ground 
state phase diagram of the 2D spin-1/2 $XXZ$ model is completed. The result 
is given in Fig.~\ref{2d} (c). There exist three different phases denoted 
by (i) $m_z=0$, (ii) $m_z\ne0$, and (iii) $m_z=0.5$ in the Figure. 
The third phase (iii) corresponds to a fully saturated ferromagnetic phase 
and is separated from the second phase (ii) by the critical 
magnetic field $h_c^{\rm max}$. This transition is trivial and is not our 
interest here. The critical magnetic field $h_c^{\rm max}$ is indeed 
easily calculated 
by going to the boson model described by Eq.~(\ref{boson}): the critical 
chemical potential to have just one particle in the $d$-dimensional boson 
system is $-2dt$ and therefore $h_c^{\rm max}/J$ is found to be $d(1+\Delta)$.

\begin{figure}[thbp]
\includegraphics[clip=true,width=0.34\textwidth,angle=0]{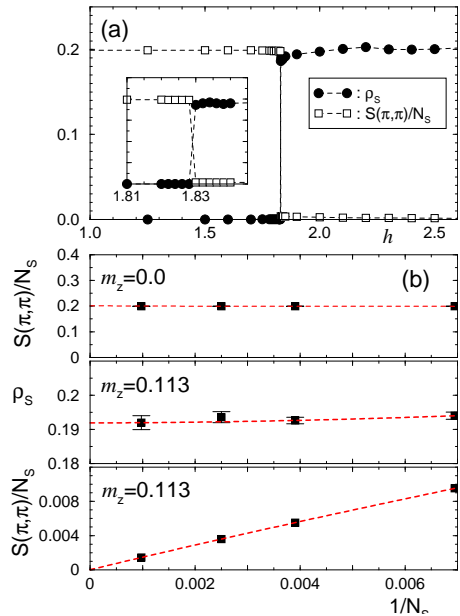}
\caption{(a) $S(\pi,\pi)$ and $\rho_s$ for the 2D spin-1/2 $XXZ$ model with 
$\Delta=1.5$ and $N_s=20^2$ as a function of the 
magnetic field $h$. Inset: enlarged scale is used. 
(b) Finite size scaling of $S(\pi,\pi)/N_s$ and $\rho_s$ for different 
magnetizations with $\Delta=1.5$. The top figure shows $S(\pi,\pi)/N_s$ 
for $m_z=0$ and the middle (bottom) figure shows $\rho_s$ ($S(\pi,\pi)/N_s$) 
for $m_z=0.113$. Dashed lines are liner fitting curves of QMC data, 
extrapolating to $1/N_s\to0$.
}
\label{sq}
\end{figure}

To show further evidence for the first-order transition between phase (i) and 
phase (ii) we calculate the spin structure factor
$S({\bf q}) = 1/N_s\sum_{{\bf i},{\bf j}}
{\rm e}^{i{\bf q}\cdot({\bf i}-{\bf j})}{\bf S_i}\cdot{\bf S_j}$ 
at ${\bf q}=(\pi,\pi)$ and the spin stiffness (helicity modulus) $\rho_s$ 
as a function of $h$ for a fixed $\Delta$. 
The spin stiffness $\rho_s$ is 
calculated 
by, for example, $\rho_s = T\langle w_x^2+w_y^2(+w_z^2)\rangle/dL^{d-2}$ 
in 2D (3D) where $w_\alpha$ is the winding number per linear spatial 
lattice size $L$ in the $\alpha$-direction~\cite{ceperley}. $\rho_s$ 
corresponds to the superfluid density in the boson model and is used to 
detect the superfluid phase in the system~\cite{mf,bat2,ceperley}. 
The results 
are shown in Fig.~\ref{sq}(a) for $\Delta=1.5$ as a function of $h$. 
When $h$ is small $S({\bf q})$ has a peak at ${\bf q}=(\pi,\pi)$ 
(${\bf q}$ dependence is not shown) and $\rho_s=0$. With increasing 
$h$ these quantities change discontinuously at $h_c$ where the 
magnetization $m_z$ jumps, and $S(\pi,\pi)$ becomes zero while $\rho_s$ 
has a finite value~\cite{com}. Apparently these two phases are 
(i) N\'eel ordered phase (for $h<h_c$) and (ii) spin-flopping phase 
(for $h>h_c$), and are separated through the 
first-order transition~\cite{note}.

Earlier studies found the supersolid phase in the boson model characterized 
by having finite values of $S(\pi,\pi)$ and $\rho_s$ 
simultaneously in a region between phase (i) and phase (ii) of the phase 
diagram~\cite{mf,bat2}. 
One can indeed barely see small but 
finite values of $S(\pi,\pi)$ for $h>h_c$ where $\rho_s$ is finite 
(see Fig.\ref{sq} (a)). 
In order to elucidate systematically the possibility of the existence of 
the supersolid phase in 2D, we carry out the finite size scaling 
analyses of $S(\pi,\pi)$ and $\rho_s$ for fixed magnetizations. 
The results are presented in Fig.~\ref{sq}(b). 
One can see that $S(\pi,\pi)/N_s$ ($\rho_s$) stays finite  in the limit 
of $N_s\to\infty$ for $h<h_c$ ($h>h_c$) while $S(\pi,\pi)/N_s$ approaches 
to zero at $N_s\to\infty$ for $h>h_c$. Doing the similar analyses 
for different 
values of $\Delta$ and $m_z$ it is found that whenever $h>h_c$ 
$S(\pi,\pi)/N_s$ goes to zero in the thermodynamic limit. 
We therefore conclude that the supersolid phase does not exist in the 
2D spin-1/2 $XXZ$ model. The results are consistent with very recent 
studies by Batrouni and Scalettar~\cite{bat}.

\begin{figure}[thbp]
\includegraphics[clip=true,width=0.34\textwidth,angle=-90]{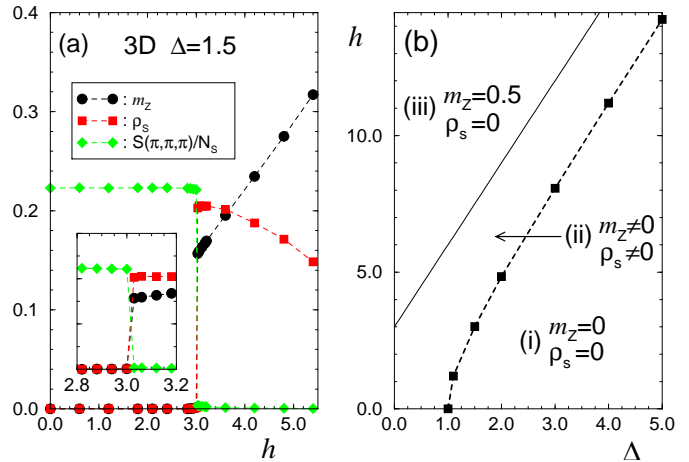}
\caption{(a) $m_z$, $S(\pi,\pi,\pi)$, and $\rho_s$ for the 3D spin-1/2 $XXZ$ 
model with $\Delta=1.5$ and $N_s=8^3$ 
as a function of the magnetic field $h$. Inset: enlarged scale is used. 
(b) The ground state phase diagram of the 3D 
spin-1/2 $XXZ$ model with anisotropic 
constant $\Delta$ in the presence of the magnetic field $h$. The diagram 
consists of (i) N\'eel ordered phase, (ii) spin-flopping phase, 
and (iii) fully saturated ferromagnetic phase. 
Solid line is $3(1+\Delta)$ (see in the text). Dashed line is a guide to 
the eye.}
\label{3d}
\end{figure}

We now carry out the similar calculations for the 3D spin-1/2 $XXZ$ model to 
study the nature of the phase transition induced by the external magnet field. 
Our strong evidence for the first-order transition is provided in 
Fig.~\ref{3d}(a). In the Figure 
magnetization $m_z$, spin structure factor $S(\pi,\pi,\pi)$, and 
spin stiffness $\rho_s$ for $\Delta=1.5$ are plotted as a function of the 
external magnetic field $h$. 
It is clearly seen that as in the case of the 2D model those quantities change 
discontinuously at the critical magnetic field $h_c$. Working with different 
values of $\Delta$, the ground state phase diagram is constructed and the 
result is shown in Fig.~\ref{3d}(b). 
As in 2D the phase diagram consists of 
three different phases: (i) N\'eel ordered phase, (ii) spin-flopping 
phase, and (iii) fully saturated ferromagnetic phase~\cite{note2}. 
The critical magnetic field $h_c$ in 3D is observed to be 
larger compared to that in 2D for a given $\Delta$. This is simply because 
the increased coordination number in 3D makes the needed magnetic field 
larger to destroy the N\'eel state. 
The similar finite size scaling analyses which was done 
for the 2D model do not find the supersolid phase in the 3D model.

\begin{figure}[thbp]
\includegraphics[clip=true,width=0.33\textwidth,angle=-90]{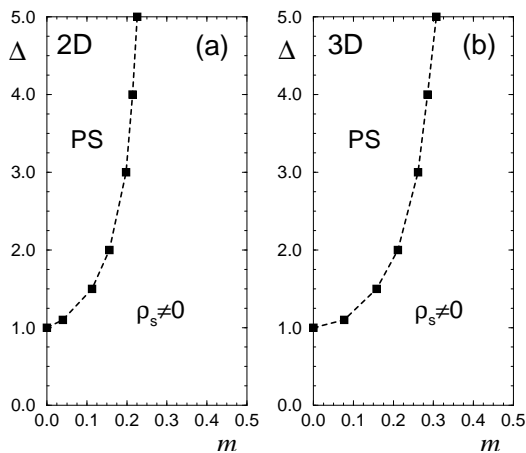}
\caption{The ground state phase diagrams of the (a) 2D and (b) 3D spin-1/2 
$XXZ$ models on the parameter space of anisotropic constant $\Delta$ 
and magnetization $m_z$. PS stands for phase separation. The N\'eel 
(fully saturated ferromagnetic) sate exists in the region of $\Delta\ge1.0$ 
and $m_z=0.0$ ($m_z=0.5$). The spin-flopping phase is in the region 
denoted by $\rho_s\ne0$. Dashed line is a guide to the eye.
For the boson model defined by Eq.~(\ref{boson}) $m_z$ and $\Delta$ 
correspond to $0.5-n$ and $V/2t$, respectively. }
\label{pd}
\end{figure}

Finally we summarize our results in Fig.~\ref{pd} by showing the ground 
state phase diagrams of the 2D and 3D spin-1/2 $XXZ$ models on the parameter 
space of anisotropic constant $\Delta$ and magnetization $m_z$.  
The phase diagrams show (i) the N\'eel ordered state at $\Delta\ge1.0$ and 
$m_z=0.0$, (ii) the spin-flopping state with $\rho_s\ne0$, 
(iii) the fully saturated ferromagnetic state at $m_z=0.5$, and (iv) a 
phase-separated state (denoted by PS in the Figures). 
The phase-separated region exists because as was seen in Figs.~\ref{2d}(a) 
and ~\ref{3d}(a) there is a region in magnetization $m_z$ to which we can 
not reach in a thermodynamically stable way no matter how fine the magnetic 
field $h$ is tuned. 
In other words, if one could have a states 
with $m_z$ which is in this magnetization region, the state would be 
phase-separated between the N\'eel state with $m_z=0$ and the spin-flopping 
state with $m_z=m_z^c$. 
Some earlier studies predicted two phase coexisting regions working with 
the canonical ensemble ({\it i.e.}, fixed $m_z$) and their results were 
interpreted as the supersolid phase~\cite{mf,bat2}. However our 
calculations conclude that these phases are thermodynamically unstable 
and are phase-separated. Our conclusions are consistent with recent studies 
by Batrouni and Scalettar for the 2D boson model~\cite{bat}.

In conclusion, we have studied numerically the magnetization process of the 
spin-1/2 anisotropic $XXZ$ model in two and three spatial dimensions using 
QMC method based on SSE and shown clear evidences of the first-order 
phase transition in the presence of an external magnetic field. 
Based on the calculated ground state phase diagrams, the existence 
of a phase-separated region and the absence of supersolid phases are 
pointed out in the related systems of hard core bosons with nearest-neighbor 
repulsions. 

It would be of great interest to study effects of random anisotropic 
constant $\Delta$ on the phase diagrams~\cite{wallin} since one can 
elucidate $quantum$ effects on recently proposed 
impurity-induced quantum-critical-point-like behavior near $first$-order 
phase transitions in the Ising models~\cite{elbio}. 
Another interesting issue to address is 
to see effects of long-ranged Coulomb repulsions between bosons on the 
phase-separated region in the phase diagram.
There is a general belief that introducing the Coulomb interactions 
replaces the phase-separated state by thermodynamically stable states 
such as a droplet-like state and perhaps a stripe state~\cite{moreo}. 
The model studied here provide an ideal system to examine the possibilities 
of those exotic states using unbiased numerical methods.


The author thanks E. Dagotto and A. Dorneich for valuable discussions.




\begin{references}
\bibitem{temp} 
Except for $\Delta\sim1.0$ for which we have to go down 
to lower temperatures to see a clear jump in the magnetization curve.
\bibitem{boson} For the corresponding boson system described by 
Eq.~(\ref{boson}), density of bosons jumps from 0.5 to $\sim0.39$ at 
critical chemical potential $\mu/t\sim2.35$, which is in good agreement 
with the results of Fig.3 in Ref.~\cite{bat} where the canonical ensemble 
was used.
\bibitem{com} In phase (iii) $\rho_s$ is zero. This is understood 
simply because the superfluid density is proportional to number density $n$ 
for the boson model and in phase (iii) $n=0$.
\bibitem{note} Except for the isotropic point at $\Delta=1$. Note also that 
there is no corresponding first-order transition for $\Delta<1$.
\end{references}
\end{document}